\documentclass[a4paper,twocolumn,prc,aps,nofootinbib,showkeys,showpacs,accepted=2024-04-01]{quantumarticle}
\pdfoutput=1
\usepackage[latin9]{inputenc}
\usepackage{color}
\usepackage{amsmath}
\usepackage{amsthm}
\usepackage{amssymb}
\usepackage{graphicx}
\usepackage{hyperref}
\usepackage[dvipsnames]{xcolor}

\makeatletter
\usepackage{dsfont}
\usepackage{epsfig}
\usepackage{amsthm}
\usepackage{tikz}
\usetikzlibrary{quantikz}
\usepackage[skins,breakable]{tcolorbox}
\tcbuselibrary{listings}
\tcbuselibrary{breakable}
\usepackage{adjustbox}

\usepackage[normalem]{ulem}

\newtcolorbox{Code}{enhanced,fonttitle=\sffamily\bfseries\large,valign=center
,drop fuzzy shadow,sidebyside,lefthand ratio=1.2,lower separated=false}

\usepackage{bm}

\usepackage{color}
\definecolor{azure}{rgb}{0.00, 0.50, 1.00}
\definecolor{darkgreen}{rgb}{0.00, 0.7, 0.30}
\definecolor{corn}{rgb}{0.80,0.5,0.10}

\makeatother

\begin{document}
\title{Efficient solution of the non-unitary time-dependent Schr\"odinger equation on a quantum computer with complex absorbing potential}

\author{Mariane Mangin-Brinet} 
\email{mariane@lpsc.in2p3.fr}
\affiliation{Laboratoire de Physique Subatomique et de Cosmologie, CNRS/IN2P3, 38026 Grenoble, France}

\author{Jing Zhang}
\email{jing.zhang@ijclab.in2p3.fr}

\affiliation{Universit\'e Paris-Saclay, CNRS/IN2P3, IJCLab, 91405 Orsay, France}

\author{Denis Lacroix }
\email{denis.lacroix@ijclab.in2p3.fr}

\affiliation{Universit\'e Paris-Saclay, CNRS/IN2P3, IJCLab, 91405 Orsay, France}

\author{Edgar Andres Ruiz Guzman}
\email{Andres.Ruiz@ibm.com}

\affiliation{Universit\'e Paris-Saclay, CNRS/IN2P3, IJCLab, 91405 Orsay, France}

\begin{abstract}
We explore the possibility of adding complex absorbing potential at the boundaries when solving the one-dimensional real-time Schr\"odinger evolution on a grid using a quantum computer with a fully quantum algorithm described on a $n$ qubit register.
Due to the complex potential, the evolution mixes real- and imaginary-time propagation and the wave function can potentially be continuously absorbed during the time propagation.   
We use the dilation quantum algorithm to treat 
the imaginary-time evolution in parallel to the real-time propagation. This method has the advantage of using only one reservoir qubit at a time, that is 
measured with a certain success probability to implement the desired imaginary-time evolution. We propose a specific prescription for the dilation method where the success probability is directly linked to the physical norm of the continuously absorbed state evolving on the mesh. We expect that the proposed prescription will have the advantage of keeping a high probability of success in most physical situations. Applications of the method are made on one-dimensional wave functions evolving on a mesh. Results obtained on a quantum computer identify with those obtained on a classical computer. We finally give a detailed discussion on the complexity of implementing the dilation matrix.  Due to the local nature of the potential, for $n$ qubits, the dilation matrix only requires $2^n$ CNOT and $2^n$ unitary rotation for each time step, whereas it would 
require of the order of $4^{n+1}$ C-NOT gates to implement it using the best known algorithm for general unitary matrices.
\end{abstract}

\maketitle

\section{Introduction}

The time evolution of quantum systems is a basic ingredient of simulating microscopic physics. It is a common subject of numerous studies, in domains ranging from 
many-body systems \cite{Smith2019,Fauseweh2020}, electron-phonon interactions \cite{Macridin2018}, to quantum field theory \cite{Jordan2012}, or even hydrodynamics \cite{Meng2023}. Most of these simulations deal with the real-time evolution of unitary systems, which fits naturally into the framework of quantum simulation. 
Quantum algorithms have also been proposed to perform    
imaginary-time evolution (ITE) or to mix real- and imaginary-time evolution on quantum computers.  Indeed, many quantum systems can be described as or reduced to systems whose norm is not conserved over time (dissipative systems, open systems, tunneling effect, instantons, \ldots), and imaginary time propagation methods are widespread in many domains. Extracting the ground state energy of microscopic systems by relying on the exponential decay of excited states is only one of the many applications of these powerful methods, which have been applied with great success not only in particle and nuclear physics but also in condensed matter and quantum chemistry. Developing algorithms on quantum computers able to efficiently encode evolution operators -- unitary or not -- is thus crucial to simulate a broad range of microscopic systems on quantum devices. Quantum algorithms aiming at including imaginary-time evolution enter either into the class 
of fully quantum algorithms or into the category of techniques referred to as hybrid quantum-classical methods \cite{Bha22}. The Quantum Imaginary-Time Evolution (QITE) method \cite{Mot19,McA19,Gom20,Lan22,Benedetti21}, which is based on the use of variational principles \cite{Yua18,End20} enters into the second class.  While expected to be less resilient to noise, fully quantum algorithms have been the subject of extensive efforts in recent years. 
Among the methods that are now being explored, one can mention the dilation method \cite{Swe15,Swe16,Spa18,Hu20,Hea21,Hu21,Tur22} allowing to treat dissipative systems, the probabilistic ITE 
\cite{Lin21,Liu21}, the ``forward and backward" real-time evolution \cite{Kos22} or the 
recently proposed Singular Value Decomposition (SVD)-based approach of Ref. \cite{Schlim2022}. 
Other methods, like the ``Linear Combination of Unitaries" \cite{Wei20,Chi12} can be used, but they 
might require a rather large number of ancillary qubits to implement the imaginary potential.

In this work, our primary motivation is to explore the simulation of the Schr\"odinger equation in position space using quantum computers, and relying completely 
on a quantum computer algorithm, i.e., avoiding  quantum-classical hybrid methods. Although solving Schr\"odinger-like equation is rather standard on classical computers, 
especially on low dimension, and despite the fact that the strategy to perform such an evolution on quantum computer 
is known from decades \cite{Bog98,Benenti2008}, the practical simulation of such problems remains, even today, rather difficult, at least due to the quantum resources required and due to the algorithm complexity (see, for instance, the discussion in \cite{Chi22}). Here we explore the possibility of adding complex absorbing potential (CAP) to a real-time 
evolution. This method is standard in classical computing to reduce the numerical resources when the evolution is solved on a grid. It consists of adding an imaginary potential 
that absorbs the wave function escaping from a certain region of interest.  
The corresponding time-dependent Schr\"odinger equation then becomes:
\begin{eqnarray}
    i \hbar \frac{\partial }{\partial t}\phi({\bf x},t) &=& \left\{ \frac{p^2}{2m} + V({\bf x}) - i W({\bf x}) \right\} \phi({\bf x},t) \label{eq:schrt}
\end{eqnarray}
that corresponds to a mixing of real- and imaginary-time propagation. 
Such potential reduces the boundary conditions effects and, when the absorption is properly made, allows for a reduction in the number of mesh points 
needed to simulate the evolution accurately. Specific discussions and technical aspects related to the implementation of CAP on classical computers can be found in for instance in Ref. \cite{Neu88,Vib92,Sei92,Ris93,Man98,Ant03,Ant08}.

A similar advantage can be envisaged when the same equation is solved on a grid on
quantum computers, when the grid points are encoded on the qubit's register. Said differently, if we can include absorbing 
boundary conditions on a quantum computer, we might significantly reduce the number of qubits needed to accurately treat the system. This reduction 
is quite attractive nowadays because the number of qubits on quantum platforms is rather restricted. In addition, efficient treatment 
of boundary conditions through absorption opens the perspective to treat certain phenomena like particle decay in real-time or scattering processes. 
A challenge in adding absorption at the boundaries is that the evolution becomes non-unitary.  

We note that the inclusion of CAP in the context of quantum 
computation has been briefly discussed in \cite{Lan22} using QITE and applied to a simple model case.
In \cite{Cha23}, although no application with CAP was shown, it was proposed to use the notion of quantum pixel for grid simulation in position. 

Here, we explore the possibility of using dilation techniques to simulate real-time evolution given by Eq. (\ref{eq:schrt}), including CAP in one dimension 
on a grid. This technique requires only one additional reservoir qubit while using fully quantum algorithms -- that is without resorting to any hybrid quantum-classical computation. The application of the imaginary evolution operator is subject to a certain success probability, which is directly related to the ``loss of norm" of the system when measuring the ancillary qubit. Special attention is given to improving the quantum algorithm to optimize the success probability. For long time evolution of systems weakly coupled to their environment or short time evolution of systems strongly coupled to their environment, the success probability of our algorithm is expected to remain close to one in many physical situations of interest. 

This article is organised as follows. 
In section \ref{Algorithm}, we detail our algorithm to simulate a non-unitary evolution given by Eq. (\ref{eq:schrt}) using the dilation method. 
Section \ref{sec:results} presents the practical implementation illustrated with results obtained on an IBM quantum simulator. The complexity associated with the implementation of the dilation method is discussed in section \ref{EffImpl}, followed by concluding remarks in section
\ref{sec:conclusion}. 

 \section{Non-unitary propagation algorithm}
 \label{Algorithm}

\subsection{Single time-step Trotter-Suzuki decomposition with imaginary time}
In order to perform the evolution given by Eq. (\ref{eq:schrt}), we first rewrite it as a generalized propagator (we take the convention $\hbar c=1$):
\begin{eqnarray}
| \Psi (t) \rangle &=& e^{-i \tilde{H} t} | \Psi (0) \rangle \label{Eq:Schr}
\end{eqnarray}
where $\tilde{H}$ is a non-unitary operator, which can be written without loss of generality as 
\begin{eqnarray}
\tilde{H}=H - i W.
\end{eqnarray}
$H= K+ V$ is the usual Hamiltonian that contains the kinetic ($K$) and potential ($V$) terms, while $W$ is an hermitian 
operator associated with the CAP. $| \Psi(0) \rangle$ is the initial state, assumed to be known and normalized to $1$. 
To implement the evolution, we consider the first time step $\Delta t$ and make use of the first order Trotter-Suzuki approximation \cite{Tro59,Suz85} and write 
\begin{eqnarray}
| \Psi(\Delta t) \rangle &=& e^{-W\Delta t}  e^{-i V \Delta t} e^{-i K \Delta t} | \Psi(0) \rangle \label{Eq:Trott}  \\
&\equiv& {\cal{U}}_{\rm approx} (\Delta t) | \Psi(0) \rangle \nonumber 
\label{eq:u_approx}
\end{eqnarray}

As in usual Trotter propagation, we can estimate the error induced by replacing (\ref{Eq:Schr}) by (\ref{Eq:Trott}). For a small time step, 
we have: 
\begin{eqnarray}
    \left\| e^{-i \tilde{H} \Delta t} - {\cal{U}}_{\rm approx} (\Delta t) \right\| &\le& \frac{\Delta t^2}{2} \left\| \left[ K, W + V \right ] \right\|  \label{Eq:Trotterror} \\
    &\le& \Delta t^2 \| K \| \left( \| W\| + \| V \| \right) \nonumber 
\end{eqnarray}
where, having in mind solving Eq. (\ref{eq:schrt}), we implicitly assumed that $[W,V]=0$, which is true if both $V$ and $W$ are 
diagonal in position space. 
This will be our case since we will consider $V$ and $W$ local in space. Here $\| \cdot \|$ denotes the spectral norm.  

The last two factors appearing in Eq. (\ref{Eq:Trott}) corresponds to unitary propagators and can usually be treated using standard techniques on 
digital computers (see illustrations in section \ref{sec:results} and discussion in appendix \ref{sec:kinetic}). Note that the kinetic-term spectral norm is explicitly given in Eq. (\ref{eq:knorm}), while $\| W\|$ and $\|V\|$ identify simply with the maximal absolute values these potentials take on the position mesh. 

Once the propagator is approximated by Eq. (\ref{Eq:Trott}), a major difficulty in implementing it on a quantum computer stems from 
the fact that the norm of the wave function immediately becomes lower than one. Indeed, for a single time step, we have 
\begin{eqnarray}
\langle \Psi(\Delta t) | \Psi(\Delta t) \rangle &=& \langle\Psi(0) | e^{-\Delta t W^\dagger} e^{- \Delta t W} | \Psi(0) \rangle \nonumber \\
&=& 1 - 2 \Delta t \langle\Psi(0) |W  | \Psi(0) \rangle + O(\Delta t^2)\nonumber 
\end{eqnarray}
where, in the last expression, we used the fact that $W$ is Hermitian. Assuming further that $W$ is positive definite will therefore lead to a decrease of the 
norm. Such a loss, which could be seen as a loss of information on the system, is impossible if the system is treated on a isolated quantum register together with a perfect quantum computers, where only unitary transformations are possible. A natural solution to this problem is to add one or several qubits that act as a reservoir and can absorb parts of the information 
on the system by interacting with it. 

\subsection{Dilation method for non-unitary propagator}
\label{sec:dilation}

\subsubsection{General description of the method}

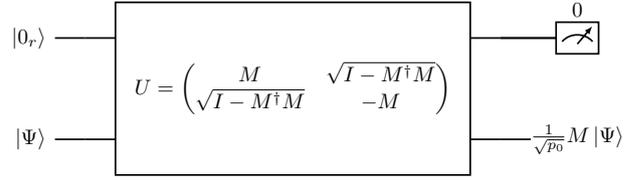
\begin{figure}
\begin{center}
\begin{adjustbox}{width=\linewidth}
\begin{quantikz}[scale=0.2]
 \lstick{\ket{0_r}} &\push{}  & \gate[wires=2][1.5cm]{U  = \begin{pmatrix} 
         M&  \sqrt{I - M^\dagger M}  \\
         \sqrt{I - M^\dagger M}              & - M
             \end{pmatrix}
 } & \push{} &\meter{0} \qw \\
 \lstick{\ket{\Psi}} & \push{} && \push{}& \frac{1}{\sqrt{p_0}} M \ket{\Psi} \qw \\
\end{quantikz}
\end{adjustbox}
\end{center}
     \caption{Circuit for the implementation of the dilation method where the reservoir+system register is prepared in the state $|0_r \rangle \otimes | \Psi \rangle$. After applying the $U$ matrix and measuring $0$ in the reservoir qubit, the system state collapses to the state given in Eq. (\ref{eq:statecollapse}). }
     \label{fig:circdil}
 \end{figure}

We briefly recall here the dilation method that is becoming today a common method for 
implementing non-unitary operators on a system encoded on a qubit register \cite{Swe15,Swe16,Spa18,Hu20,Hea21,Hu21,Tur22}. 
Let us assume that we want to implement $| \Psi \rangle \rightarrow M | \Psi \rangle $ where $M$ 
is not unitary, but is assumed to be hermitian and satisfying $ \left\| M \right\|\le 1$. We assume the system is encoded on $n$ qubits, leading to a computational basis of size $2^n$. 
The dilation method relies on doubling the size of the computational basis by adding a single reservoir qubit and applying 
the following unitary matrix to the $n+1$ qubits:
 \begin{eqnarray*}
U=\begin{pmatrix} 
         M&  \sqrt{I - M^\dagger M}  \\
         \sqrt{I - M^\dagger M}              & - M
             \end{pmatrix} . \label{eq:Uenlarged}
\end{eqnarray*} 
Specifically, we have 
\begin{eqnarray}
    U \left(|0_r \rangle \otimes | \Psi \rangle \right) &=&   |0_r \rangle \otimes \left[ M |\Psi \rangle \right] %
    \nonumber \\    &+&  
     |1_r \rangle \otimes  \left[ \sqrt{I - M^\dagger M}  |\Psi \rangle \right], \nonumber 
\end{eqnarray}
where $\{ |{0}_r \rangle , |{1}_r \rangle \}$ denote the two states associated to the reservoir qubit while $| \Psi \rangle$ is the initial 
system state. From the above relation, we see that we can perform the desired operation by preparing the initial state $|0_r \rangle \otimes | \Psi \rangle$, applying 
the $U$ matrix and measuring $0$ in the reservoir state. The procedure is schematically represented in the circuit displayed in Fig. \ref{fig:circdil}. 
 More precisely, when measuring the reservoir qubit in state $|0\rangle$, according to the Born measurement's rules, the resulting reservoir+system state is given by:
 \begin{eqnarray}
     | \Psi_0 \rangle &=& \frac{1}{\sqrt{p_s}} | 0_r \rangle \otimes \left[ M | \Psi \rangle\right] \label{eq:statecollapse}
 \end{eqnarray}
$p_s$ is nothing but the probability to measure the reservoir qubit in state $|0 \rangle$, called hereafter simply success probability and given by:
\begin{eqnarray}
    p_s &=& \langle \Psi | M^\dagger M | \Psi \rangle.  \label{eq:success} 
\end{eqnarray}

\subsubsection{Application to imaginary-time propagation}

We use the dilation method to apply the CAP, i.e., $M \propto e^{- W \Delta t}$. A similar problem was addressed in Ref. \cite{Tur22} where the dilation method was implemented to perform imaginary-time propagation using directly the full Hamiltonian, i.e., assuming $W=H$, 
to obtain the ground state of the problem. In that case, it was proposed to use the prescription 
\begin{eqnarray}
    M &=& \frac{1}{\sqrt{I + e^{-2W \Delta t}}} e^{- W \Delta t}. 
\end{eqnarray}  
We first implemented this prescription but realized that it has the clear drawback that the success probability 
exponentially tends to $0$ when performing several time steps. This could indeed easily be seen, considering that for arbitrary time-step $\Delta t$, we always have $p_s \le 1/2$ provided that $M$ is positive definite. If the dilation process is iterated $r$ times to simulate the evolution 
up to $t= r \Delta t$, the success probability will be lower than $1/2^{r}$. This aspect is critical for practical implementation since, after several steps, most measurements of the ancillary qubit will be rejected, and the 
method rapidly becomes inefficient. 

This unwanted feature can easily be avoided by taking the simpler prescription $M = e^{-W\Delta t}$ leading 
to  
\begin{eqnarray}
U=\begin{pmatrix} 
         e^{-  W \Delta t}&  \sqrt{I - e^{-2 W \Delta t}} \\
         \sqrt{I - e^{-2    W \Delta t}}              & -e^{- W \Delta t} 
\end{pmatrix}. \label{Umodified}
\end{eqnarray} 
The unitarity of the matrix $U$ can be easily proven noting that $[\sqrt{I - e^{-2 W \Delta t}},e^{- W \Delta t}]=0$. 

Applying this prescription to a given initial state $|\Psi_{\rm ini}\rangle$, we see that
the success probability given by Eq. (\ref{eq:success}) becomes:
\begin{eqnarray}
    p_s (\Delta t) &=& \langle \Psi_{\rm ini} | e^{-2 W \Delta t} | \Psi_{\rm ini} \rangle \nonumber \\
    &\simeq&   \langle \Psi_{\rm ini} | \Psi_{\rm ini} \rangle - 2 \Delta t \langle \Psi_{\rm ini} |W | \Psi_{\rm ini} \rangle +  O(\Delta t)^2,
    \nonumber 
\end{eqnarray}
and provided the initial state is normalized to $1$ at initial time, the success probability remains close to $1$ if $\Delta t$ is small, which is much more favorable than the prescription of Ref. \cite{Tur22}. 

In terms of the specific steps taken for time evolution Eq. (\ref{eq:schrt}), 
one can take advantage of the Trotter decomposition(\ref{Eq:Trott}) as follows:
\begin{itemize}
    \item For the first time step, the propagators $e^{-i V \Delta t} e^{-i K \Delta t}$ are directly implemented as a set of unitary 
gates on the system register.
    \item The CAP term is then implemented using the dilation method by measuring the reservoir qubit in state $|0\rangle$.  After the measurement, the wave function, denoted hereafter as $| \Psi_{\rm dil} (t) \rangle$, is given by:
    \begin{eqnarray}
    | \Psi_{\rm dil} (\Delta t) \rangle &=& \frac{1}{\sqrt{p_s(1)}}  {\cal{U}}_{\rm approx} (\Delta t) |\Psi (0) \rangle 
    \end{eqnarray}
    where ${\cal{U}}_{\mathrm{approx}}$ is the evolution matrix from Eq. \eqref{eq:u_approx}.
 We denote here by $p_s(1)$ (resp. $p_s (r)$) the success probability associated with the first (resp. the $r^{th}$) 
    application of the dilation method. 
    \item Iterating the above procedure, after the $r^{th}$ application, and denoting $t= r \Delta t$, we deduce that 
    the system wave function is given on the qubit register as:
    \begin{eqnarray}
        | \Psi_{\rm dil} (t) \rangle = \frac{1}{\sqrt{{\cal P}_s(t)}}  \left[ {\cal{U}}_{\rm approx} (\Delta t)\right]^r |\Psi (0) \rangle,  
        \label{eq:psi_dil}
    \end{eqnarray}
    where the global probability of success ${\cal P}_s(t=r\Delta t)$ relates to $r$ consecutive successful performances of the non-unitary evolution with CAP, i.e. the probability of obtaining $r$ times zero when measuring the reservoir qubit. This probability is given by 
    \begin{eqnarray}
    {\cal P}_s(t=r\Delta t) &=& p_s(1) \cdots p_s(r).  \label{eq:success1r}
    \end{eqnarray}
\end{itemize}

\subsubsection{Physical estimates of the total success probability evolution}
\label{sec:psp}

\begin{figure}[htbp]
  \centering
  \includegraphics[width=0.4\textwidth]{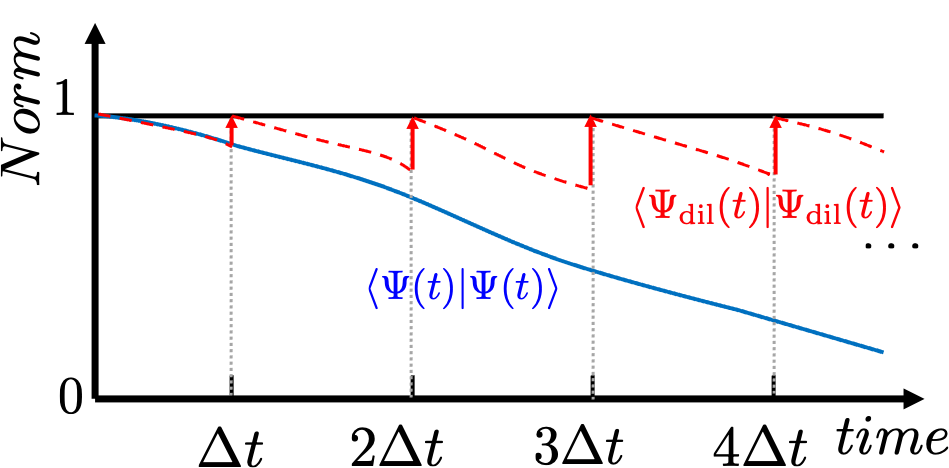}
  \caption{Schematic view of the norm evolution when solving the Schr\"odinger equation given by Eq. (\ref{eq:schrt}) on a classical computer 
  (blue solid line). The horizontal black line and vertical gray dotted lines are added as guidance. The red dashed line corresponds to the evolution of
  the norm of the system state encoded on the qubit register when the problem is solved using the dilation method with the prescription 
  $M = e^{-W \Delta t}$. At each time step $\Delta t$, when the reservoir qubit is measured in state $|0\rangle$, 
  the system state is renormalized to have a norm equal to $1$. 
  }
  \label{fig:normschem}
\end{figure}

A schematic representation of the iterative procedure is depicted in Fig. \ref{fig:normschem}, where the norm of the state 
$| \Psi_{\rm dil} ( t) \rangle$ is shown as a function of time. Each time a zero is measured in the reservoir qubit, 
the norm is reinitialized to 1. 

Denoting by $| \Psi(t) \rangle$ the wave function obtained with good accuracy by solving numerically the Schr\"odinger equation (\ref{eq:schrt})
on a classical computer, according to Eq. (\ref{Eq:Trotterror}), 
we see that we have:
\begin{eqnarray}
    \left[ {\cal{U}}_{\rm approx} (\Delta t)\right]^r |\Psi (0) \rangle \sim  | \Psi(t) \rangle
\end{eqnarray}
within the accumulated Trotter errors. 
Consequently, we also have:
\begin{eqnarray}
   \lim_{\Delta t \to 0}  {\cal P}_s(t)    =  \langle \Psi(t) | \Psi(t) \rangle \equiv {\cal N}(t) . \label{eq:success2}
\end{eqnarray}
The last property makes the dilation method rather attractive for physical systems simulation. Indeed, we have shown that the success probability at a given time $t$ tends, in the limit $\Delta t\to 0$, to the norm of the wave function 
that survives to the absorption when solving the problem on a classical computer.
In many physical situations where CAP is useful, we are interested in systems that are 
rather localized in a certain region of space and where part of the wave function is emitted. This is, for instance, what happens when particles 
are emitted from a compact localized quantum object. In this case, most particles are emitted over certain time-scale $t  \le \tau_{\rm decay}$,  implying 
that ${\cal N}(t)$ decreases and then reaches a stationary asymptotic value. Accordingly, we expect in such situation that the global probability of success ${\cal P}_s(t>\tau_{\rm decay})$ defined in equation (\ref{eq:psi_dil}) remains constant, which implies that single-step success probability $p_s(r) \simeq 1$ for $r \gg \tau_{\rm decay}/\Delta t $. In brief, for such 
systems where most particles are emitted after a certain transient time, almost all events will be successful after this time. 
Note that, in section \ref{sec:results}, we will actually consider the free wave-packet propagation, the physical situation where all particles escape the grid at infinite time. This case corresponds to an extreme example of decay phenomena, where the importance of absorption is more pronounced since ${\cal N}(t)$ tends to $0$ at infinite time. 
It is therefore a perfect example for testing the absorption of particles at the grid boundaries.   

\subsubsection{Expectation values of non-absorbed observables}

In general, we are often interested in computing the expectation values of observables, generically denoted by $O$ 
taken on a state $| \Psi (t)\rangle$. Provided that we have $| \Psi_{\rm dil} ( t) \rangle$ encoded on the system register, 
we can compute approximately the quantity:
\begin{eqnarray}
 \langle O \rangle_{\rm dil} &=& \langle   \Psi_{\rm dil} ( t) | O |  \Psi_{\rm dil} ( t) \rangle 
\end{eqnarray}
by implementing either a Hadamard test \cite{NielsenChuang} and performing a set of measurements, or, if the observable is written as a linear 
combination of Pauli chains, one can directly measure the system register after a proper change of the measurement basis 
to estimate each individual Pauli chain (see, for instance, \cite{Ayr23}, table 3). 

In parallel, we can also estimate the total success probability, denoted by ${\cal P}'_{s}(t)$ simply by counting the number of events where only zeros were 
measured in the reservoir qubit and divide this number by the total number of events. 
From these estimates, we can compute approximate expectation values on $| \Psi(t)\rangle$ simply as $ {\cal P}'_{s}(t) \langle O \rangle_{\rm dil} $.

\section{Illustration: 1D Schr\"odinger evolution on a grid with complex absorbing potential}
 \label{sec:results}

We give here a proof of principle of applying the strategy discussed above. 
Specifically, we show examples of 1D Schr\"odinger equation on a grid with absorbing boundary 
conditions. 
 We first concentrate on one critical situation where $V(x)=0$. In this case, assuming 
that the particle is located inside the grid and will freely spread during the evolution, it should eventually
completely be absorbed by the CAP in the infinite time limit. 
We then apply our method to the case of a wavepacket in a 
gaussian potential, where part of the wavefunction remains trapped
in the potential over a long period of time. All calculations have been made below using the Qiskit software 
that emulates a perfect digital computer \cite{Qis21}.  

\subsection{Free wave propagation with CAP}

We consider a problem described on a restricted area of space $x \in [x_{\rm min}, x_{\rm max}]$. The wave function 
is assumed to be initially a Gaussian wave function located in the middle of the region with:
\begin{eqnarray}
\Psi(x,t=0) = \frac{1}{\pi^{1/4} \sigma^{1/2}} e^{\left[-\frac{(x-x_0)^2}{2\sigma^2} + i\frac{m.v}{\hbar}.(x-x_0)\right]} ,\label{eq:initial}
\end{eqnarray}
with $x_0=(x_{\rm max}+x_{\rm min})/2$. Eventually, we will also consider the possibility that the wave function 
has an initial boost proportional to the parameter $v$. In the simulation, we use natural units with the convention $\hbar=c=1$, $\hbar^2/2m =1$. 

We concentrate here on the free wave case, i.e., we assume $V(x) = 0$ in Eq. (\ref{eq:schrt}). Our goal is to remove 
the wave function while approaching the boundaries, i.e., to apply an absorbing potential within a certain distance $D$ such that either $|x-x_{\rm min} |< D$ or $| x_{\rm max} - x| < D$.    

We consider here as absorbing potential $W$, the local amplitude operator suggested by Kosloff et al. in \cite{Kos1986}:
\begin{eqnarray}
    \left\{
    \begin{array}{lll}
    W(x) = & \displaystyle \frac{U_0}{\cosh^2[\alpha.(x_{\rm max} - x)]}, & {\rm for}\; D > x_{\rm max}  - x   \\
    \\
    W(x) = & \displaystyle \frac{U_0}{\cosh^2[\alpha.(x-x_{\rm min})]}, & {\rm for}\; x-x_{\rm min}  < D  \\
    \\
    W(x) = & \displaystyle 0, & {\rm everywhere~else} . 
    \end{array}
    \right. \nonumber 
\end{eqnarray}
The ``amplitude reduction" technique described \cite{Kos1986} 
consists in applying the usual evolution operator to propagate from 
$|\Psi(t) \rangle$ to  $|\Psi(t+\Delta t)\rangle$ and consecutively applying on $|\Psi(t+\Delta t)\rangle$ an ``absorption"
step according to  
\begin{eqnarray}
|\tilde{\Psi}(t+\Delta t)\rangle= (1-W dt)|\Psi(t+\Delta t)\rangle. \label{AmpRed}
\end{eqnarray}
This method is equivalent to implementing an absorbing potential at the boundaries using a first-order Trotter-Suzuki decomposition. This justifies the use of the amplitude operator $W$ proposed in \cite{Kos1986} in our simulation, even though we do not use amplitude reduction technique. 

To solve the problem on a classical or quantum computer, we directly discretize 
the one-dimensional space on a grid of equally spaced mesh points. The wave function $|\Psi\rangle$ is then written as:  
\begin{eqnarray}
|\Psi\rangle=\sum_{i=0}^{N_x-1} \psi(x_i)| x_i \rangle ~~ \mbox{with} ~~ \psi(x_i)=\langle x_i |\Psi \rangle. 
\label{InitialState}
\end{eqnarray}
The space grid $\{ x_i =x_{\rm min}+ i . \Delta x \}_{i=0,N_x -1}$ ranges from $x_{\rm min}$ to 
$ x_{\rm max}$, with a mesh step given by $\Delta x = (x_{\rm max}-x_{\rm min})/(N_x-1)$. On the discretized mesh, the CAP becomes
\begin{eqnarray}
W(x_i)&=&\frac{U_0}{\cosh^2[\alpha\Delta x.(N_x-1-i)]}, \quad {\rm for}\;\; i>N_x-k \nonumber  \\
W(x_i)&=&\frac{U_0}{\cosh^2[\alpha \Delta x . i]},\quad \quad \qquad \qquad {\rm for}\;\; i<k \nonumber
\end{eqnarray}
where $k = D/\Delta x$ is a fixed integer. 

As a reference calculation, we first solve the problem on a classical computer using the split-operator method \cite{Fei82,Bal97}, i.e., going back-and-forth from 
position to momentum space. Consistently with the quantum computation case (see below), we used the simplest first-order splitting. In the classical computer case, the absorbing potential is directly applied to the wave function by multiplying each component $\psi(x_i)$ by $e^{-\Delta t W(x_i)}$. In this case, the norm of the wave function decreases in time, as shown in Fig. \ref{fig:normschem} directly probing the effect of the absorption at the two boundaries of the mesh. Such classical computing results are rather standard and will serve as reference 
calculations for the one obtained using the quantum computing algorithm.

\begin{figure}[htbp]
  \centering
  \includegraphics[width=0.49\textwidth]{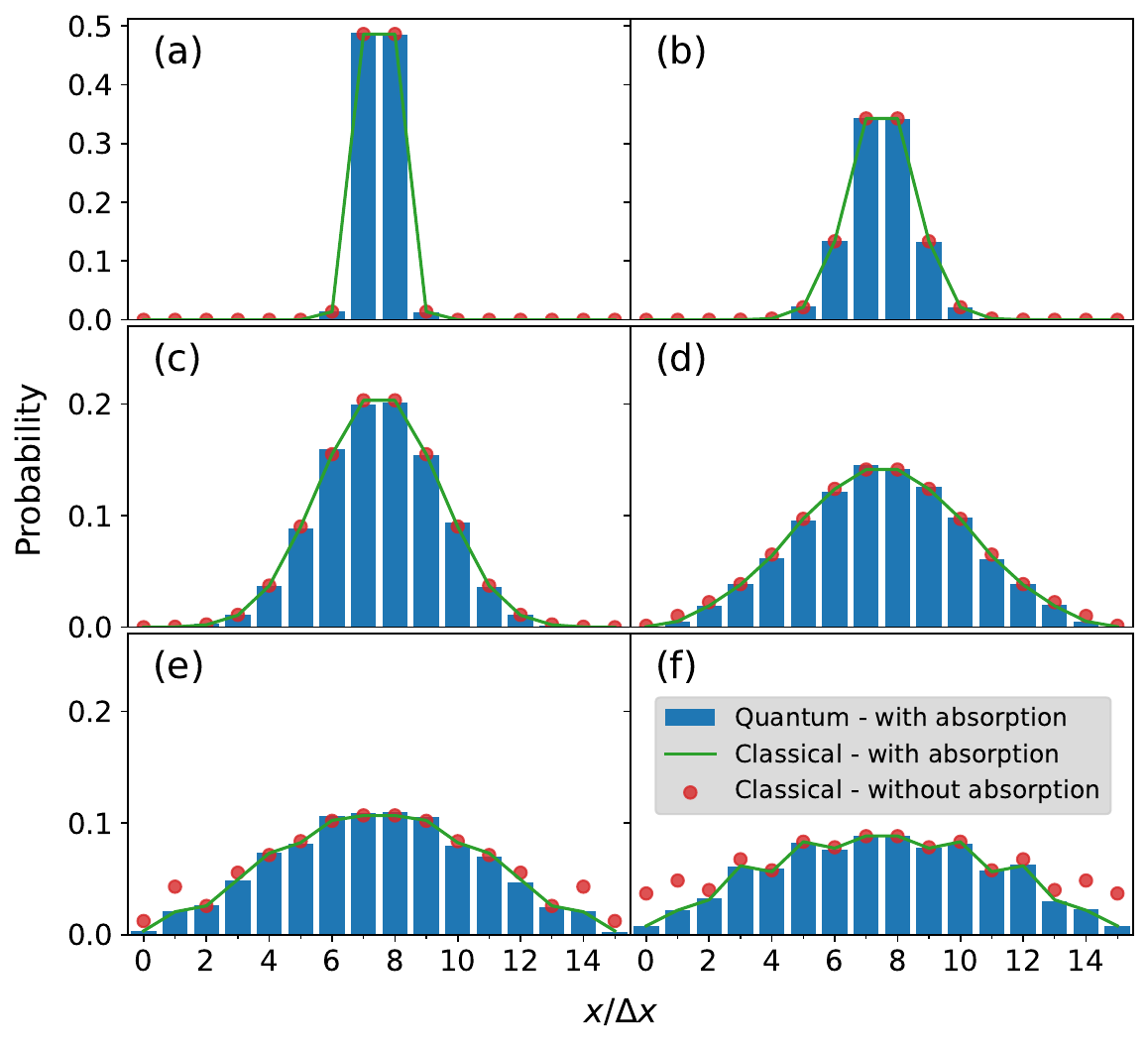}
  \caption{Time evolution of a wave-packet with no boost. The initial wave packet is shown in panel (a). Its half-width is chosen to be 
  $\sigma = 0.4$. Snapshots of the wave function amplitudes at time $t=r\Delta t$ where $r=1, 2, \cdots, 5$ are shown in panel (b) to (f) respectively with a time step $\Delta t = 1.2$. The parameters $U_0=0.4$ and $\alpha = 1.5$
  have been used for the absorption potential. In the quantum simulation case, for each time step, the amplitudes are reconstructed from the measurement of the register 
  system qubit using $2^{14}$ shots. For such a large number of shots and in the absence of device noise, the error bars are essentially zero. 
  Note that the y-axis is zoomed in panels (c-f).  }
  \label{fig:CftAbs1}
\end{figure}

\begin{figure}[htbp]
  \centering
  \includegraphics[width=0.49\textwidth]{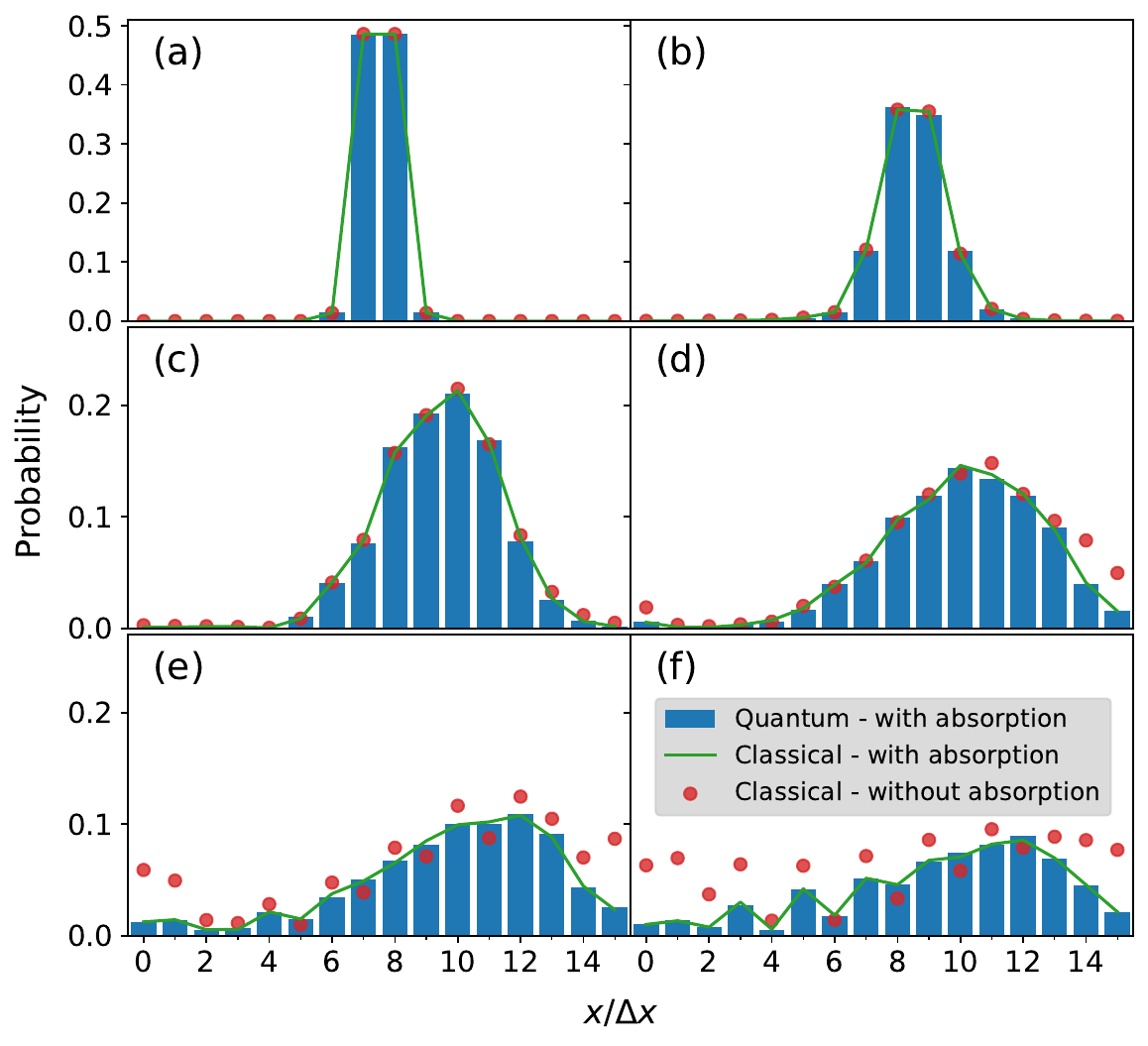}
  \caption{Same as Fig.~\ref{fig:CftAbs1} but with a boost corresponding to $v=4$.}
  \label{fig:CftAbs2}
\end{figure}

We show in Fig. \ref{fig:CftAbs1} and Fig. \ref{fig:CftAbs2} two examples of wave function evolution without and with an initial boost respectively, obtained with a classical computer. 
Results have been simulated using an initially localized state located in the middle of the mesh with a width $\sigma=0.4$  
and $N_x = 2^4=16$ points. The CAP parameters are $U_0=0.4$ and $\alpha=1.5$. For figure \ref{fig:CftAbs2}, the boost parameter value is set to $v= 4$. 
In these figures, we display both the results of the calculations with and without absorption. In the absence of absorption, we see the accumulation of the wave function amplitude that is reflected by the boundaries and then starts to interfere with the incoming wave function. 
As it is well known from classical computing, such interference prevents from properly describing the wave function spreading. 
This interference is reduced when the CAP is included, although some interference is still visible, especially in the boosted 
case. Note that such interference can be reduced by fine-tuning the absorption potential, but this is not the aim of the present work. 
Our main objective is to give proof that the same evolution as obtained in a classical computer can be implemented on quantum computers using the dilation method  discussed above.

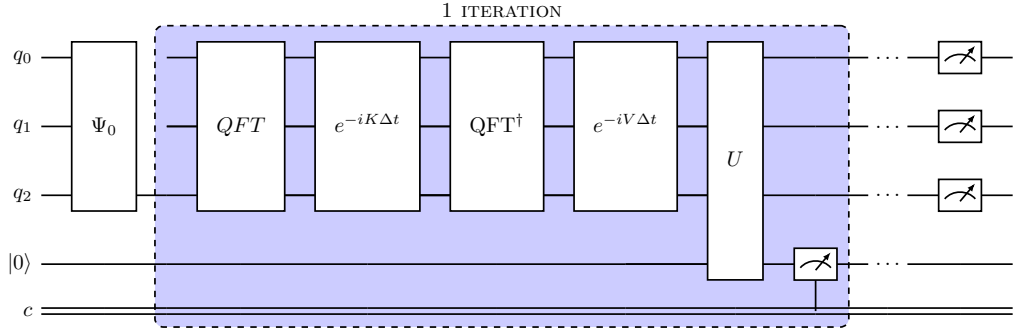
\begin{figure*}[htbp]
\begin{centering}
\begin{tikzpicture}
\node[scale=0.8] {
\begin{quantikz}
\lstick{$q_0$} & \gate[3,disable auto height]{\begin{array}{c} \text{} \\ \text{$\Psi_0$} \\ \text{} \end{array}} &    \gategroup[wires=5,steps=7,style={dashed,
                   rounded corners,fill=blue!20, inner xsep=2pt},
                   background]{{\sc 1 iteration}}
 &  \gate[3,disable auto height]{\begin{array}{c} \text{ } \\ \text{$QFT$} \\ \text{} \end{array}}  &
  \gate[3,disable auto height]{\begin{array}{c} \text{ } \\ \text{$e^{-iK\Delta t}$} \\ \text{} \end{array}}  &
 \gate[3,disable auto height]{\begin{array}{c} \text{ } \\ \text{QFT$^\dagger$} \\ \text{} \end{array}}  &
   \gate[3,disable auto height]{\begin{array}{c} \text{ } \\ \text{$e^{-iV\Delta t}$} \\ \text{} \end{array}}  &
    \gate[4,disable auto height]{\begin{array}{c} \text{} \\ \text{$U$} \\ \text{} \end{array}} &
\qw&\ \ldots\ \qw &\meter{} &\qw\\
\lstick{$q_1$} &  &   & \qw & \qw&\qw&\qw &\qw& \qw&\ \ldots\ \qw &\meter{}  &\qw\\
\lstick{$q_2$} &  &   \qw& &\qw&\qw\qw&\qw&\qw& \qw&\ \ldots\ \qw& \meter{} &\qw\\
\lstick{$\ket{0}$} &   \qw&\qw&\qw& \qw& \qw & \qw& \qw & \meter{} \vqw{1}  &\ \ldots\ \qw&\qw &\qw\\
\lstick{$c$} &   \cw&  \cw& \cw&\cw&\cw& \cw & \cw & \cw& \cw&\cw &\cw
\end{quantikz}
};
\end{tikzpicture}\\ 
\par\end{centering}
\begin{centering}
\caption{Schematic illustration of one time-step evolution including the implementation of the CAP using the reservoir qubit as discussed in section \ref{sec:dilation} together with the first order Trotter-Suzuki method where the QFT is used to go back and forth from momentum 
to position space. After a certain number of time step iterations, the probability density is obtained
by measuring the qubits used to describe the system. Note that in the specific schematic example shown here, there are only 3 such qubits.
}
\label{fig:propquant}
\par\end{centering}
\end{figure*}

\subsection{1D free wave evolution with CAP on a quantum computer}

To encode the problem onto the qubit register, we use the standard binary (SB) representation that consists of mapping each position $|x_i \rangle$ into 
the register state $| [i] \rangle$ where $[i]$ denotes the binary representation of the integer $i$. 

The first-order Trotter-Suzuki method is used to perform the time evolution. Note that at first order, Trotter-Suzuki decomposition is strictly equivalent for the evolution to the ``amplitude reduction" technique implemented on the classical computer discussed previously. As in the classical 
computer simulation, we perform the evolution of the Eq. (\ref{eq:schrt}) by going back and forth from the position space 
where the operators $(V,W)$ are diagonal, and the associated propagator can be easily implemented  
to the momentum space, where the same holds for the kinetic term $K$. 
Changing from one basis to the other requires the use of the Quantum Fourier Transform (QFT) algorithm or its inverse (see appendix \ref{sec:kinetic}). 
This gives the scheme depicted schematically in Fig. \ref{fig:propquant}.

The absorbing potential is included using the dilation method introduced in section \ref{sec:dilation}. 
The blue box shown in Fig. \ref{fig:propquant} is repeated $N_t$ times, 
each time the result of the ancilla measurement being stored in a classical register, and the total system is measured only at the end. We 
keep only runs for which all the measurements of the reservoir qubit lead to zero.

The $U$ matrix is implemented using the algorithm
discussed in section \ref{EffImpl}. 
For a given time, the amplitude is reconstructed by measuring the system register. We used here $2^{14}$ shots for each panels of 
Fig. \ref{fig:CftAbs1} and \ref{fig:CftAbs2}.
We see in both figures that wave function amplitudes obtained using the quantum algorithm 
perfectly match the ones obtained using the classical computer algorithm. Note that we have systematically compared 
the results obtained on classical and quantum processor units, including or not an initial boost and/or adding 
a local potential $V(x)$, and always obtained a perfect matching in the results.

\begin{figure}[htbp]
  \centering
  \includegraphics[width=0.9\linewidth]{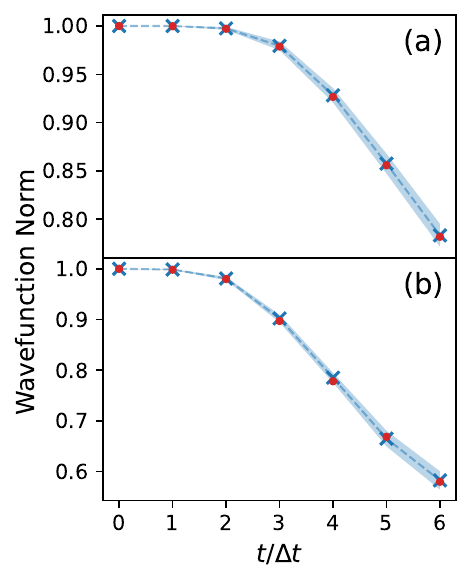} 
  \caption{Evolution of the norm of the wave function decreasing in time due to the CAP for the 
  case of a freely expanding wave function without (a) and with boost (b). In each panel, the classical 
  computer simulation (red filled circle) is compared to the one obtained on the quantum computer with the dilation method (blue cross -- dashed line). In the latter case, we directly show the success probability (see Eq. (\ref{eq:success2})) that is equal to the wave function norm when the number of measurements tends to infinity. 
  Here, $2^{10}= 1024$ measurements are used for each point obtained in the quantum simulation. 
  The errorbars, displayed by the dashed area, correspond to the standard deviation obtained by repeating 20 runs, each with $2^{10}$ events 
  and by estimating the width around the mean value within the 20 runs. Note that using $2^{14}$ events instead of $2^{10}$ as is used in 
  Fig. \ref{fig:CftAbs1} and \ref{fig:CftAbs2}
 leads essentially to vanishing error bars.}
  \label{fig:norm}
\end{figure}

\subsection{Success probability and absorption}
\label{sec:successprob}

The good agreement between the quantum and classical computer simulations is further confirmed in Fig. 
\ref{fig:norm} where we compare the evolution of the wave function norm during the time evolution for the two cases shown in Fig. \ref{fig:CftAbs1} and \ref{fig:CftAbs2}. In the classical computer, this norm can be directly computed by integrating the wave function over the grid. 
As discussed in detail in section \ref{sec:psp} (see also Fig. \ref{fig:normschem}), in the quantum calculation case, 
the system wave function is re-normalized to $1$ after each measurement of the reservoir qubit. To 
compute the physical norm, i.e., the one corresponding to the non-absorbed particle, we can use its connection 
with the success probability given by Eq. (\ref{eq:success2}). In practice, the success probability at a given time $t = r \Delta t$ is given by the probability to only obtain $0s$ in the measurements of the reservoir 
qubit for time steps $1$ to $r$. The norm shown in the quantum calculation is deduced, assuming a strict equality 
between the success probability and the norm of the wave function. We see that the two calculations are almost on top of each other. 
As a side remark, we also implemented the prescription of Ref. \cite{Tur22}. 
As we discussed in the 
introduction, in this case, the success probability decays exponentially with the number of times 
the dilation is implemented. For comparison, without boost, 
the success probability is of the order of $1/2^r$ after the $r^{th}$ time step, which should be compared to the values reported in Fig. \ref{fig:norm}.

\subsection{Wave propagation with a confining potential}

In the previous section, we tested the methodology in the extreme situation where the wave function 
expands and will eventually disappear completely from the mesh. Accordingly, the probability of success in our case will rapidly tend to zero, rendering the approach rather difficult to use due to the high rejection rate. Most of the physical applications where the CAP is employed, however, will lead
asymptotically to cases where part of the wave function remains trapped inside the numerical mesh. In contrast, the remaining part is emitted from the localized source. This is 
,for instance, the case when a system is excited and emits particles. Then, the fraction of 
emitted particles will depend on the internal excitation and particle emission threshold. 
In these physical situations, an essential aspect of the method we process is that the total success probability will not asymptotically tend to zero but saturate to the fractions of particles remaining in the numerical mesh. This is illustrated 
here by adding a confining potential.

\begin{figure}[htbp]
  \centering
  \includegraphics[width=1.0\linewidth,height=5.5cm]{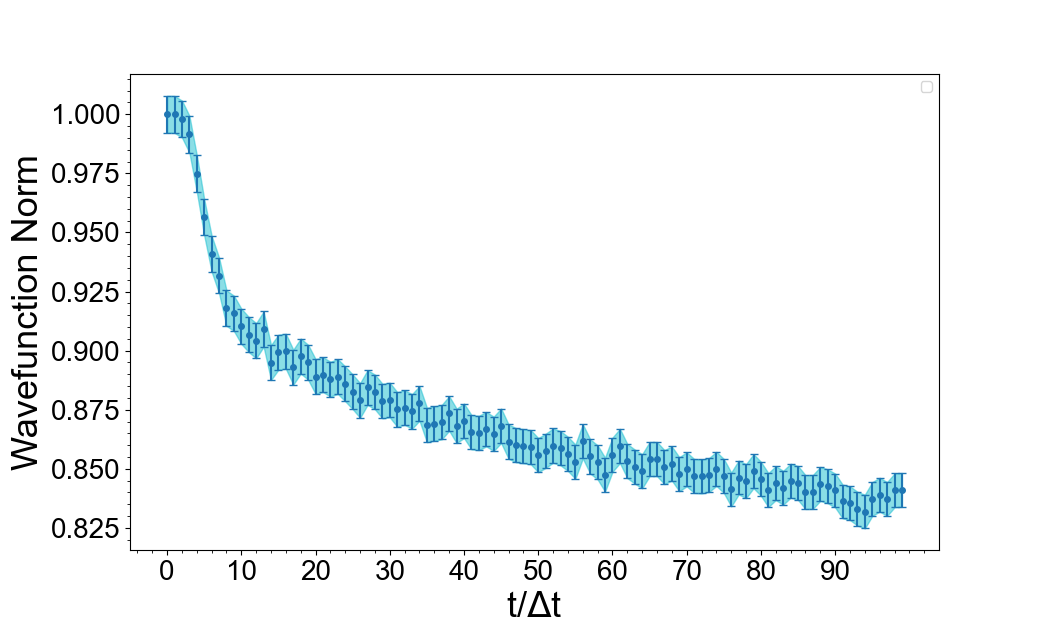} 
  \caption{Evolution of the norm of the wave function in the presence of a Gaussian potential. The shaded area represents the statistical error bars. The norm is constructed directly from the success probability and we used $2^14$ shots.}
  \label{fig:normpot}
\end{figure}

Specifically, in addition to the CAP, we now add a Gaussian fixed potential centered in the integration domain and given by
\begin{eqnarray*}
V(x)=V_0 \displaystyle e^{\frac{x^2}{2\sigma^2_V}}
\end{eqnarray*}
with $V_0=-1$ and $\sigma_V=1$. The evolution of the norm of the wave-function -- which is the success probability of measuring the reservoir qubit in state $|0\rangle$, see Eq. (\ref{eq:success1r}) --  is displayed in Figure \ref{fig:normpot}, over 100 time steps. We can clearly see on this plot the advantage of our method, which is to lead to asymptotically non-vanishing total probability of success since it is directly related to the probability that the system remains in the box. 

\section{Complexity analysis for the implementation of the dilation matrix}
\label{EffImpl}

The dilation algorithm we use to implement non-unitary propagation
is expected to be one of the most efficient algorithms in terms of ancillary qubits since it requires the addition of only one reservoir qubit. 
Still, it requires implementing a general matrix $U$ which is quite demanding (see, for instance, Ref. \cite{Kro22} for a comprehensive overview). We restrict the discussion here on the complexity in terms of gates for implementing this matrix directly without adding more ancillary qubits besides the one needed in the dilation.  

If $n$ qubits are used to simulate the physical system, it leads to a matrix $U$ of dimension $2^{n+1}\times 2^{n+1}=2^d\times 2^d$, whose naive implementation, as a general unitary matrix, 
requires in principle of the order of $d^2 \times 4^d$ gates \cite{NielsenChuang}. 
In \cite{Sch22}, it was shown that, as an alternative to the dilation technique, 
the general complexity of implementing a matrix treating the imaginary-time propagation can be reduced by a factor 2 when using the singular value decomposition (SVD) technique. 
We show here that, due to the specific diagonal nature of the absorbing potential, and using the Cosine-Sine decomposition, it is possible to implement the dilation matrix $U$ with much lower number of gates than the SVD-based approach. 

To decompose arbitrary unitary operators of dimension $2^{d}\times2^{d}$ in terms of C-NOT and one-qubit gates, one of 
the most efficient decomposition scheme is the Quantum Shannon Decomposition (QSD) \cite{She06,Kro22}.
This algorithm is based on the 
Cosine-Sine Decomposition (CSD) \cite{Tuc99} and requires $(3/4)\times4^d-(3/2)\times2^d$ C-NOT gates, 
where $d$ is the total qubits number. 
If we use the CSD algorithm alone, without the improvements brought by
the Shannon decomposition, we get a total complexity of 
$4^d-2^{d+1}$ C-NOT gates and $4^d$ elementary one-qubit gates
\cite{Mottonen2004}. 
QSD and CSD algorithms are particularly well-suited to our case 
since the dilation matrix can immediately be written
in terms of cosine-sine decomposition:
\begin{eqnarray}
U&=&
             \begin{pmatrix}
             \mathds{1} & 0 \\
             0  &   -\mathds{1}
              \end{pmatrix}
       \begin{pmatrix} 
         e^{- \tau W}&  \sqrt{I - e^{-2 \tau W}} \\
        - \sqrt{I - e^{-2 \tau W}}              & e^{- \tau W} 
           \end{pmatrix} , \nonumber\\
         &=& \begin{pmatrix}
             \mathds{1} & 0 \\
             0  &   -\mathds{1}
              \end{pmatrix}
       \begin{pmatrix} 
         C& S\\
       -S    & C
        \end{pmatrix} , \label{CSD}
\end{eqnarray} 
where $C$ and $S$ matrices verify $C^2+S^2=\mathds{1}$. 
We note that, due to the particular structure of the dilation matrix, it can always be written directly in this form. 

The l.h.s. matrix in (\ref{CSD}) is a trivial multiplexor that can be implemented with one $Z$ gate acting on qubit $d$. However, since we are only interested in the diagonal part of the dilation matrix to implement the non-unitary evolution, we can even remove this $Z$ gate and consider directly the non-trivial part of the decomposition, namely the matrix $D=       \begin{pmatrix} 
         C& S\\
       -S    & C
        \end{pmatrix}
$ \footnote{Note also that, without lost of generality the  multiplexor can also directly be avoided in the first place simply by redefining $U$ as:
\begin{eqnarray}
U=\begin{pmatrix} 
         e^{-  W \Delta t}&  \sqrt{I - e^{-2 W \Delta t}} \\
         -\sqrt{I - e^{-2    W \Delta t}}              & e^{- W \Delta t} 
\end{pmatrix}, \label{Ualternative}
\end{eqnarray} instead of Eq. (\ref{Umodified}). }.

Thanks to the particular form
of the dilation matrix in the case of non-unitary propagation with absorbing potential (see Eq. (\ref{Umodified})), we can even further reduce this complexity. 
Our matrices $C$ and $S$ are indeed diagonal, and we only need to resort to a small part of the CSD algorithm to implement $U$. 
These matrices can be written as 
\begin{eqnarray}
    C = {\rm diag}[\cos(\theta_i)], ~ S={\rm diag}[\sin(\theta_i)] \nonumber 
\end{eqnarray}
with $i=0,\ldots 2^{n}-1$. For a given time $\tau$, the angles are related to the imaginary potential through:
\begin{eqnarray}
e^{-W_{ii}\tau}=\cos{\theta_i}, ~\sqrt{1-e^{-2W_{ii}\tau}}=\sin{\theta_i}\label{Eq:angles}
\end{eqnarray}
where we can restrict $\theta_i$ angles to $[0,\frac{\pi}{2}]$. 

\begin{figure}[htbp]
\begin{center}
\includegraphics[width=\linewidth]{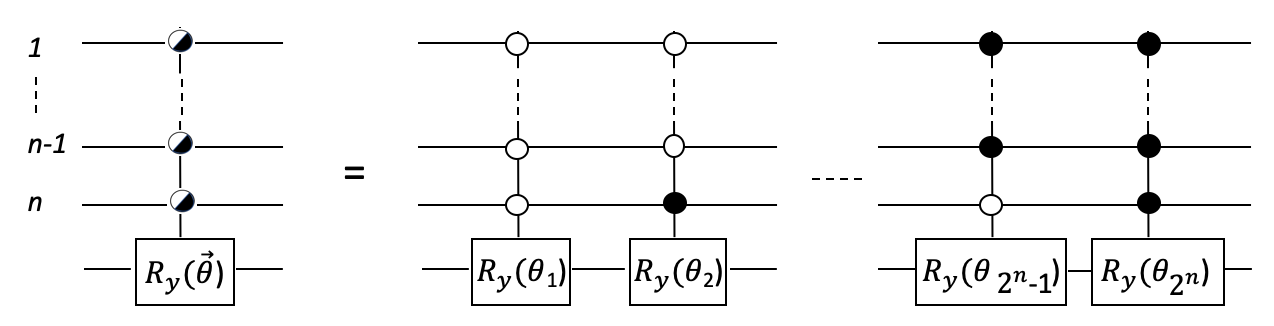}  
\caption{Definition of an $n$-fold uniformly controlled rotation on qubit $n+1$, $F_{n+1}^{n}(R_{y}(\vec{\theta})$ (see text). The black control bits stand for value 1 and the white for 0. This operator match the $D$ matrix we use to implement the non-unitary propagation.}\label{Fig:UnifCont}
\end{center}
\end{figure}

The operator $D$ is a uniformly controlled rotation about the $y$ axis, also denoted as $F_{n+1}^{n}(R_{y}(\vec{\theta}))$ \cite{Mottonen2004}. It consists of $n$-fold controlled rotations of qubit $n+1$ about the
axis $y$, one $R_y$ rotation for each of the $2^{n}$ different classical values of the control qubits. The circuit representation of $D=F_{n+1}^{n}(R_{y}(\vec{\theta}))$ is displayed 
in Figure \ref{Fig:UnifCont}. In general, $F_{n+1}^{n}(R_{y})$ is a product of $2^{n}$ two-level operators. In \cite{Mottonen2004} is proposed an implementation of $F_{n+1}^{n}(R_{y})$ 
in $2^{n}$ C-NOT and $2^{n}$ rotations acting on qubit $n+1$. 

\begin{figure}[htbp]
\begin{center}
\includegraphics[width=\linewidth]{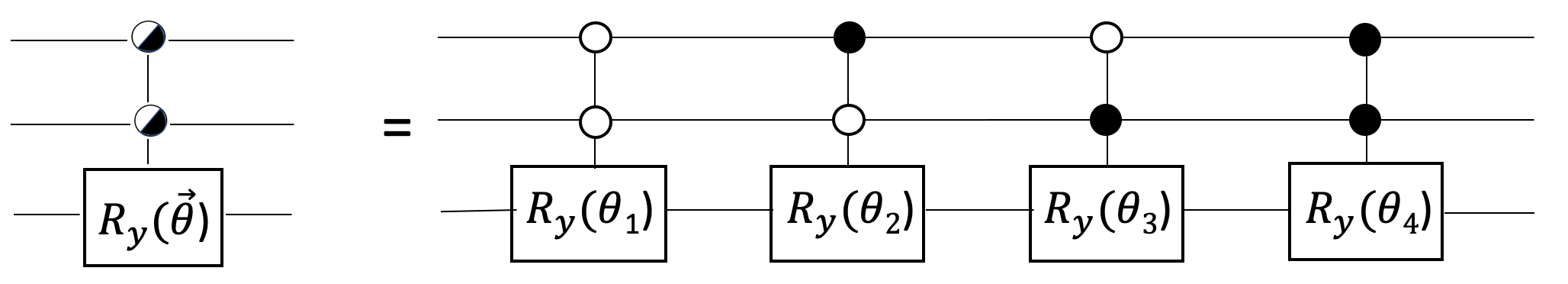} 
\caption{{\it{ Definition of the uniformely controlled rotation $F_3^2(\vec{\theta})$. The angles $\theta_i$ are defined in equation (\ref{Eq:angles}).}}}\label{Fig:F32}
\end{center}
\end{figure}

\begin{figure}[htbp]
\begin{center}
\includegraphics[width=\linewidth]{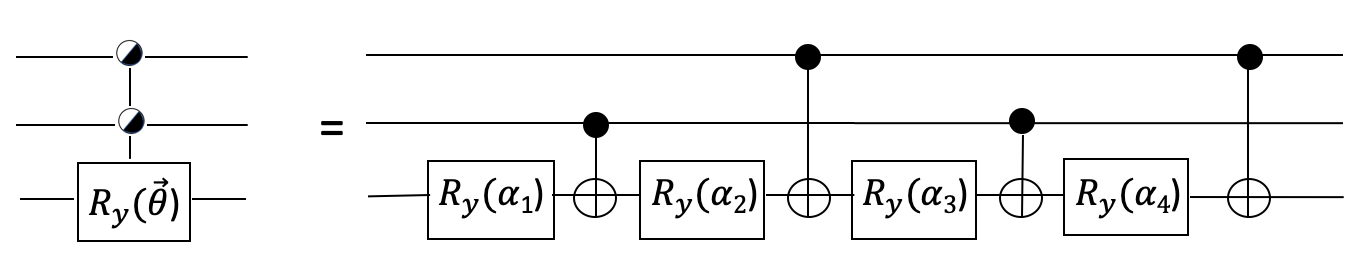} 
\caption{{Efficient implementation of uniformly controlled rotation, using $2^{n}$ C-NOT and $2^{n}$ one-qubit gates. Example for $F_3^2(\vec{\theta})$, showing the $2^2=4$ rotation gates and 4 C-NOT needed.}}\label{Fig:EffUnifCont}
\end{center}
\end{figure}

We illustrate the efficient implementation of the uniformly controlled
rotation on the $n=2$ case. To fix notations, the definition of $F_3^2(\vec{\theta})$ is displayed in Figure \ref{Fig:F32}. 
\begin{table*}
\begin{tabular}{l| cccccl }
  \hline
  qubit number ~~$d=(n+1)$   &  ~(2+1)~ & ~(3+1)~ & ~(4+1)~ & ~(5+1)~&   $\cdots$ & $(n+1)$ \\
  \hline 
  CSD (optimized) \cite{Mot06} &  26& 118 & 494 &  2014 & $\cdots$ & $\frac{1}{2} (4^{n+1} -2^{n+1}) - 2 $  \\
  QSD (optimized) \cite{She06} & 20& 100 & 444 & 1868 & $\cdots$ & $\frac{23}{48} 4^{n+1} - \frac{3}{2} 2^{n+1} - 2 $ \\
  \hline
  \hline 
  SVD (for CAP)  \cite{Schlim2022}   &  6 & 14 & 30 & 62 &$\cdots$ & $2^{n+1} - 2$  \\
 Dilation(QSD for CAP) [this work]   &  4 & 8 & 16 & 32 &$\cdots$ &  $2^{n}$
\end{tabular}
\caption{\label{tab:complexity} Illustration of the number of CNOT required to implement a unitary matrix of dimension $d\times d$ 
with $d=(n+1)$. The two first lines are adapted from table I of Ref. \cite{Kro22} and concerns general unitary matrices. The two last lines consider the specific case 
where the CAP using local complex potential is implemented using either the SVD-based method of \cite{Schlim2022}, or the dilation method we propose to use here. }
\end{table*}
An illustration of an efficient circuit for $F_{3}^{2}(R_{y}(\vec{\theta}))$, as proposed in Ref. \cite{Mottonen2004}, 
 is shown in Figure \ref{Fig:EffUnifCont}. 
The rotation angles $\alpha$ used in the 
efficient circuit are linked to the original rotation angles $\theta$ by a linear transformation. 

In the specific case we consider here, i.e., with diagonal CAP, the implementation of the dilation method thus requires 
$2^n$ C-NOT and $2^n$ one-qubit gates. For comparison, in the same situation considered here, the SVD-based matrix requires solely to 
implement the diagonal matrix $\Sigma$ introduced in \cite{Schlim2022} of size $d \times d$ and will require $2^{n+1} - 1$ $z$-rotation 
unary gate and $2^{n+1} - 2$ CNOT gates. We compare in table \ref{tab:complexity} the number of required CNOT 
for some optimized methods as a function of the system size register $n$. We see that the dilation method outperforms the other techniques for the specific implementation of local complex potential. For instance, the numerical simulation considered in Fig. \ref{fig:CftAbs1} and \ref{fig:CftAbs2} have been made using $n=4$ qubits, for which a direct implementation of the $U$ matrix requires 
only 16 CNOT and 16 $R_y$ operations per time-step.

\section{Conclusion}\label{sec:conclusion}

We analyze the possibility of performing wave function Schr\"odinger evolution on a grid 
where the evolution mixes both real- and imaginary-time evolution. 
The physical motivation behind adding an imaginary potential is the possibility 
to absorb particles that might be emitted from a localized source without the need 
to treat very extended space regions. This technique is standard in classical computing and 
might potentially lead to a significant reduction in the number of qubits when the grid itself
is encoded on the qubit register. 

For the real-time evolution, we use the standard first-order Trotter-Suzuki method, where the system is 
evolved alternatively in position and momentum space. We show that absorbing boundary conditions can be 
efficiently implemented using the dilation method. In particular, we propose a specific prescription 
for the ingredient of the dilation method that ensures that the success probability remains significant 
in most situations due to its connection with the physical wave function norm. 

Proof of the principle of the technique is given in the case of a one-dimensional particle evolving freely 
on a mesh and being continuously absorbed in the two mesh sides. We show that results obtained on a 
quantum computer are identical to those obtained on a quantum computer and that, due to the local nature of the 
absorbing potential usually used in applications. Finally, it is demonstrated that the dilation matrix can be implemented efficiently, significantly reducing the required quantum resources.   

The solution of Schr\"odinger equation with absorbing boundary conditions is an archetype of an example where
real-time and imaginary time propagation are mixed. Developing a specific strategy to implement such a mixing   
serves us as a test bench to test efficient quantum algorithms. However, when the number of available qubits will increase 
the accessible dimension of meshes might be large enough so that imaginary potential might not be necessary. Nevertheless, we would like to mention that this exploratory study and the resulting algorithms might be exported to more complex problems in the near future, like configuration interaction techniques involving complex energy due to the presence of resonances \cite{Mic21} that are very hard 
to solve in large Hilbert spaces on classical computers. 

\subsection{Acknowledgments }
This project has received financial support from the CNRS through
the 80Prime program  and the AIQI-IN2P3 project. J. Zhang is funded by the joint doctoral programme of Universit\'e Paris-Saclay and the Chinese Scholarship Council. This work is part of 
HQI initiative (www.hqi.fr) and is supported by France 2030 under the French 
National Research Agency award number "ANR-22-PNQC-0002".
We acknowledge
the use of IBM Q cloud as well as the use of the Qiskit software package
\cite{Qis21} for performing the quantum simulations. The Qiskit code used for the quantum simulation presented in this work is available on request from the authors.

\appendix 

\section{Practical aspects of the kinetic term implementation}
\label{sec:kinetic}

For completeness, we give below some details regarding the implementation of the kinetic term in momentum space (see also \cite{Benenti2008}). 
Since $K$ is diagonal in momentum representation, we use the Fourier transform to go from
space to momentum representation, apply $e^{-i\frac{p^2}{2m}\Delta t}$, and then revert back to position representation. 
If $F$ denotes the Fourier transform operator, we have:
\begin{eqnarray*}
e^{-iK\Delta t} = F^{\dagger}e^{-i\frac{p^2}{2m}\Delta t}F.
\end{eqnarray*}
So to apply the operator $e^{-iK \Delta t}$, we first perform 
a standard QFT, whose general expression for a 
standard basis state $| x\rangle$ is \cite{NielsenChuang}
\begin{eqnarray*}
| x\rangle  \longrightarrow \frac{1}{2^{n/2}} \bigotimes_{l=1}^{n} \Big[ |0\rangle +e^{\frac{2i\pi x}{2^l}} |1\rangle\Big].
\end{eqnarray*}
Once in Fourier space, we must implement the diagonal operator $e^{-ip^2\Delta t/2m}$ in the basis $| p\rangle$. 
Our qubit basis is defined in terms of $N=2^n$ discretized space points, $n$ being the number of qubits considered, such that $x_k=x_{\rm min}+k.\Delta x $ where $x_{\rm min} \le x \le x_{\rm max}$, $\Delta x =\frac{x_{\rm max}-x_{\rm min}}{2^n-1}=\frac{L}{2^n-1}$ is the discretization step in position space. 
The discretized momentum $p$ can be written as : 
$p=(\frac{2\pi}{L})\sum_{j=0}^{n-1} 2^j p_j$, with $p_j=\{0,1\}$ and $p\in [0;\frac{2\pi}{L}(2^{n}-1)]$. 
However, in this case, we have no negative values for the momentum. To adapt our Fourier transform to a wave packet with both negative and positive momenta, we shift the $p$ range to 
$p=\frac{2\pi}{L}\Big(\sum_{j=0}^{n-1} 2^j p_j-2^{n-1}$\Big). Then, for the range of momentum, we have:
\begin{eqnarray}
    p\in \left[-\frac{2\pi}{L}2^{n-1};\frac{2\pi}{L}(2^{n-1}-1)\right] \label{eq:interval}
\end{eqnarray}
\begin{eqnarray*}
p_k=\frac{2\pi}{L} \Big(1-\frac{1}{2^n} \Big)\Big(\sum_{j=0}^{n-1} 2^j k_j-2^{n-1}\Big) \qquad \mbox{$k_j\in\{0,1\}$}
\end{eqnarray*}
 
Then, we deduce:  
\begin{eqnarray*}
p^2&=&\Big(\frac{2\pi}{L}\Big)^2 \Big(1-\frac{1}{2^n} \Big)^2\Big(\sum_{j=0}^{n-1}2^{2j}k_j+\sum_{j=0}^{n-1}\sum_{l>j}^{n-1}2^{l+j+1}k_lk_j\\
&-&\sum_{j=0}^{n-1} 2^{n+j} k_j+2^{2n-2}\Big)
\end{eqnarray*}

So we have
\begin{eqnarray*}
e^{-i\frac{H}{\hbar}\Delta t} |p_{n-1}\ldots p_0\rangle&=&\\
\bigotimes_{j=0}^{n-1} e^{-i\frac{(2^{2j}-2^{n+j})p_j}{2m\hbar}\Delta t}
&\displaystyle{\Pi_{k=j+1}^{n-1}}&e^{-i\frac{2^{k+j+1}p_kp_j}{2m\hbar}\Delta t} |p_j\rangle
\end{eqnarray*}
We dropped the last term since it corresponds to  a global phase on the qubits. To implement this evolution, we thus need 
only phase gates and controlled-phase gates. 

An inverse Fourier transform brings us back to the direct space, where we can implement the part of the evolution 
operator containing the potential.

Finally, we mention  that the above treatment of the momentum space leads to the kinetic term spectral  norm (see the interval given in (\ref{eq:interval})):
\begin{eqnarray}
    \| K \| = \frac{1}{2m} \|p\|^2 = \frac{\pi^2}{L^2 m }2^{2n-1}. \label{eq:knorm}
\end{eqnarray}

\end{document}